\documentclass[aps,prc,twocolumn,showpacs,floatfix,nofootinbib,preprintnumbers,superscriptaddress]{revtex4-1}

\usepackage{CJKutf8}
\usepackage{epsfig}
\usepackage{color}
\usepackage{rotating}
\usepackage{natbib}
\usepackage{verbatim}
\usepackage{epstopdf}
\usepackage{graphicx}
\usepackage{changes}
\usepackage{bm}
\usepackage{hyperref}
\usepackage{multirow}
\usepackage{makecell}

\usepackage{enumerate}
\usepackage{amsmath}
\usepackage[normalem]{ulem}

\newcommand{\soft}[1]{\texttt{#1 }}
\newcommand{\lisepp}{\soft{LISE$^{++}$}}

\begin{document}

\begin{CJK*}{UTF8}{gbsn}

\title{Beyond the proton drip line:  Bayesian analysis of proton-emitting nuclei}

\author{L{\'e}o Neufcourt}
\affiliation{Department of Statistics and Probability, Michigan State University, East Lansing, Michigan 48824, USA}
\affiliation{Facility for Rare Isotope Beams, Michigan State University, East Lansing, Michigan 48824, USA}

\author{Yuchen Cao (曹宇晨)}
\affiliation{Facility for Rare Isotope Beams, Michigan State University, East Lansing, Michigan 48824, USA}
\affiliation{National Superconducting Cyclotron Laboratory, Michigan State University, East Lansing, Michigan 48824, USA}

\author{Samuel Giuliani}
\affiliation{Facility for Rare Isotope Beams, Michigan State University, East Lansing, Michigan 48824, USA}
\affiliation{National Superconducting Cyclotron Laboratory, Michigan State University, East Lansing, Michigan 48824, USA}

\author{Witold Nazarewicz}
\affiliation{Facility for Rare Isotope Beams, Michigan State University, East Lansing, Michigan 48824, USA}
\affiliation{Department of Physics and Astronomy, Michigan State University, East Lansing, Michigan 48824, USA}

\author{Erik Olsen}
\affiliation{Institut d'Astronomie et d'Astrophysique, 
Universit{\'e} Libre de Bruxelles, 1050 Brussels, Belgium}

\author{Oleg B. Tarasov}
\affiliation{National Superconducting Cyclotron Laboratory, Michigan State University, East Lansing, Michigan 48824, USA}

\date{\today}

\begin{abstract}
\begin{description}

\item[Background] 
The limits of the nuclear landscape are determined by nuclear binding energies. 
Beyond the proton drip lines, where the  separation energy becomes negative, there is not  enough binding energy to prevent protons from escaping the nucleus. Predicting properties of unstable nuclear states in the vast territory of proton emitters  poses an appreciable  challenge for nuclear theory as it often involves  far extrapolations.
In addition, significant discrepancies between nuclear models in the proton-rich territory call for quantified predictions.

\item[Purpose]
With the help of Bayesian methodology, we mix a family of nuclear mass models corrected with statistical emulators trained on the experimental mass measurements, in the proton-rich region of the nuclear chart.

\item[Methods]
Separation energies were  computed within nuclear density functional theory using several Skyrme and Gogny energy density functionals. We also considered mass predictions based on two models  used in astrophysical studies.
Quantified predictions were  obtained for each model using Bayesian Gaussian processes trained on separation-energy residuals and combined via Bayesian model averaging.

\item[Results]
We obtained a good agreement between  averaged predictions of statistically corrected models  and experiment.  In particular, we  quantified model results for one- and two-proton separation energies and derived probabilities of proton emission. This information enabled us to produce a quantified landscape of proton-rich nuclei. The most promising candidates for  two-proton decay studies have been identified.

\item[Conclusions]
The methodology used in this work has broad applications to model-based extrapolations of various nuclear observables. It also provides a reliable  uncertainty quantification  of theoretical predictions.
\end{description}
\end{abstract}

\maketitle
\end{CJK*}

\section{Introduction} 

Much of the nuclear landscape of bound nuclides remains unexplored \cite{Erl12a,Agbemava2014}. The one- and two-proton drip lines lie relatively close to the line of beta stability  due to the presence of the Coulomb barrier that has a confining effect on the proton density. As a result,  relatively long-lived,
 proton-unstable nuclei can exist beyond the  drip line~\cite{Pfutzner2004,Blank2008,Pfutzner12,Pfutzner13,Giovinazzo2013}. The vast territory of proton-unstable nuclides contains rich and unique  information on nuclear structure and dynamics  in the presence of the low-lying proton continuum. 

Of particular interest is the phenomenon of ground state two-proton (2$p$) radioactivity  found in a few very proton-rich even-$Z$ isotopes,  in which  single proton decay is energetically forbidden or  suppressed due to proton
 pairing and the resulting odd-even  binding energy effect \cite{Goldansky1960}. 
Currently, 2$p$ radioactivity has been detected in a handful of nuclei: $^{19}$Mg~\cite{Mukha2007}, 
$^{45}$Fe~\cite{Giovinazzo2002,Pfutzner2002}, $^{48}$Ni~\cite{Dossat2005,Dossat2007,Pomorski2011,Pomorski2014},  $^{54}$Zn~\cite{Blank2005,Ascher2011},
and  $^{67}$Kr  \cite{Goigoux2016}. In addition, several broad resonances associated with 2$p$ decay were reported in, e.g., $^6$Be \cite{Egorova12} and $^{11,12}$O \cite{Webb19,Webb19b}. The unique experimental data on lifetimes and correlations between emitted protons has
triggered considerable theoretical interest \cite{Golubkova16,Grigorenko2017,Gigorenko2018,Kostyleva2019,Oishi17,SiminKr67,Wang19}.

The positions of particle decay thresholds are determined by the nuclear binding energy through measured masses. In the regions where experimental mass data are absent, nuclear models must be deployed to provide the missing information about the topography of the mass surface. In this context,  the quality of theoretical mass predictions  can be significantly improved when aided by the current experimental information through machine learning techniques \cite{Athanassopoulos2004,Bayram2017,Yuan2016,Utama16,Utama17,Utama18,Bertsch2017,Zhang2017,Niu2018,Rodriguez2019,Rodriguez2019a}. Recently, we developed the statistical framework of Bayesian Gaussian process techniques to quantify patterns of systematic deviations between theory and experiment by providing statistical corrections to average prediction values, and develop full uncertainty quantification on predictions through credibility intervals \cite{Neufcourt2018}. The quantified predictions of individual models enabled us to carry out 
Bayesian model averaging (BMA)  analysis of nuclear masses \cite{Neufcourt2019}. In this way, the ``collective wisdom" of several relevant models could be maximized by providing the best prediction rooted in the most current experimental information.

In this paper, using several variants of BMA,  
we quantify the predicted binding-energy surface in the region of $2p$ radioactivity. To this end,  we employ several global mass models to determine
the posterior probability  for each proton-rich nucleus to exhibit ground-state proton or $2p$ emission. We find that extrapolations for proton drip-line locations are fairly consistent across the mass models used, in spite of significant variations between their raw predictions. In this respect, this study should be considered as an extension of the previous work \cite{Olsen2p,Olsen13}, in which the most promising candidates for $2p$ emitters were identified by considering several mass models. Here, we limit our investigations to nuclei with $Z\le 82$ as  it is predicted  \cite{Olsen2p,Olsen13} that above lead  the $\alpha$-decay mode  dominates and no measurable candidates for $2p$ emission are expected.

The paper is organized as follows. Section~\ref{MassModels} presents the global nuclear mass models used in our study.  The statistical methodology is described in Sec.~\ref{StatModels}. The results obtained in this study are discussed in Sec. \ref{Results}. Finally, Sec.~\ref{Conclusions} contains a summary and conclusions.

\section{Nuclear mass models}\label{MassModels} 

Our global mass calculations are based on nuclear density functional theory (DFT) with
several energy density functionals (EDFs). As in the previous studies \cite{Neufcourt2018,Neufcourt2019}, we considered the Skyrme functionals
SkM$^*$ \cite{Bartel1982}, SkP \cite{Dob84}, SLy4 \cite{Chabanat1995}, SV-min \cite{Kluepfel2009}, UNEDF0 \cite{UNEDF0},  UNEDF1 \cite{UNEDF1}, and  UNEDF2 \cite{UNEDF2}.
In this work, we have enriched the set of nuclear models with two additional EDFs: D1M and BCPM. The functional
D1M~\cite{Goriely09} is a modern parametrization of the finite-range Gogny
interaction, optimized to  2149 measured masses from the 2003
mass evaluation (AME2003) \cite{AME03b}, charge radii, and  nuclear matter properties.
In the functional BCPM \cite{Baldo13}, the bulk part of the functional is given by a
fit to the microscopic EOS in both neutron and symmetric nuclear matter.
This formulation of the functional  results in 
a relatively small number of free
parameters that are adjusted to reproduce the experimental binding energies of
579 even-even nuclei of AME2003. 
 
For each Skyrme EDF, the mass table of even-even nuclei was computed self-consistently by solving the Hartree-Fock-Bogoliubov (HFB) equations. Binding energies of odd-$A$ and odd-odd nuclei were obtained from the binding energy values and average pairing gaps computed for even-even neighbors. In this respect, this work follows closely the methodology described in Refs.~\cite{Erl12a,Olsen13,Neufcourt2018,Neufcourt2019}.
For D1M and BCPM, the binding energies of odd-$A$ and odd-odd nuclei are
computed by solving the HFB equations for one- and two-quasiparticle configurations
with the appropriate constraint on particle
number~\cite{Duguet2001}.
The above set  of DFT models was augmented by two mass  models commonly used in nuclear astrophysics studies:  FRDM-2012 \cite{Moller2012} and HFB-24 \cite{Goriely2013}. 

It is to be noted that while the proton chemical potential is positive for proton unbound nuclei, the HFB results obtained with the discretized continuum are very stable in the considered range of binding energies. This is because the Coulomb barrier tends to confine the proton density in the nuclear interior and effectively pushes the continuum up in energy \cite{Dobaczewski94,Vertse00} on the proton-rich side. 
Consequently, the proton separation energies $S_{1p}$ and  $S_{2p}$ and the corresponding $Q$-values can be  obtained safely  from the predicted binding energies.

The candidates for the true $2p$ decay were
selected according to the energy criterion used in the global survey \cite{Olsen2p}:
\begin{equation} \label{eq:2p}
Q_{2p} > 0,~S_{1p} > 0.
\end{equation}
This condition
corresponds to  the simultaneous
emission of two protons as the sequential emission of two
protons is energetically forbidden.

\section{Statistical methods}\label{StatModels}

Our Bayesian methodology for building Gaussian process (GP)  emulators to produce quantified extrapolations of theoretical nuclear model predictions beyond the experimental data range has been extensively developed 
in our previous work  \cite{Neufcourt2018,Neufcourt2019}. Here, we incorporate two statistical innovations, a non-zero GP mean parameter and a new Bayesian calculation of  model mixing weights.
 
\subsection{Gaussian process}
The Bayesian statistical model for the separation-energy residuals (i.e., differences between experimental and theoretical values) $y_i= y^{exp}(x_i) - y^{th}(x_i)$ can be written as:
\begin{equation}
y_{i}= f(x_i, \theta)+\sigma\epsilon_{i},
\end{equation}
where the function $f(x,\theta)$ represents the systematic deviations 
and $\sigma \epsilon$ the statistical uncertainty 
propagated from the uncertainty on the statistical model parameters.

Quantified extrapolations $y^*$   are obtained from the posterior distribution $p(y^*|y)$ using a stationary Markov chain.
Similarly to our previous studies, we model independently $S_{1p}$ and $S_{2p}$ on the  subsets of nuclei defined by particle-number parities (even-even, even-odd, etc.).
By doing this we are ignoring some (slight) correlations 
between systematic uncertainties.

For the function $f$ we take a Gaussian process, i.e., a Gaussian functional on the two-dimensional nuclear domain indexed by $x=(Z, N)$, characterized by its mean $\mu$ and covariance $k$:
\begin{equation}\label{GPmodel}
f(x,\theta)\sim\mathcal{GP}(\mu, k_{\eta,\rho}(x, x')).
\end{equation}
We have found in a previous study \cite{Neufcourt2018} 
that Gaussian processes overall outperform Bayesian neural networks, achieving similar root-mean-squared (rms) deviations 
with a more faithful uncertainty quantification and considerably fewer parameters.
In our previous studies,
we took for simplicity the GP mean to be uniformly zero. 
Here we take it as an additional scalar parameter.
The results below show that this can improve the rms deviation by an
additional $15\%$, compared to the initial $25\%$ refinement brought by the GP.
In order to model the ``spatial"
dependence of nearby nuclei, 
we use an exponential quadratic covariance kernel: 
\begin{equation}\label{kernel}
k_{\eta ,\rho }(x,x^{\prime }):=\eta ^{2}e^{-\frac{(Z-Z^{\prime })^{2}}{%
2\rho _{Z}^{2}}-\frac{(N-N^{\prime })^{2}}{2\rho _{N}^{2}}},
\end{equation}
where the parameters $\theta \equiv \{\eta ,\rho_{Z},\rho_{N}\}$ 
have a straightforward interpretation: 
$\eta $ defines the scale  and $\rho_{Z}$ and $\rho_{N}$
are characteristic  correlation ranges  in the proton and neutron directions,
respectively.
Consequently,  our statistical model has four parameters
 $\theta:=(\mu,\eta,\rho_Z,\rho_N)$. 
 
Samples from posterior distributions were obtained 
from 50,000 iterations of Monte Carlo Markov Chains (MCMC), 
after the stationary state was reached 
(50,000 samples are previously burnt-in), 
which were used in turn to generate 10,000 mass tables. 
Each series of simulations required 16,128 core hours distributed on 
96 machines on computing clusters.

\subsection{Datasets}

Our dataset combines all experimental masses from
AME2003 \cite{AME03b} and AME2016+, which contains the AME2016 dataset \cite{AME16b} augmented by masses from Refs.~\cite{deRoubin17,Welker17,JYFLTRAP,Leistenschneider2018,Michimasa2018,Oxford18,Ito18}.
For nuclei where experiments have been repeated, 
we keep the most recent value.
For testing purposes we split this dataset into a training set (AME2003) 
and a testing set 
(AME16-03: all masses in AME2016+ that are not in AME2003).
For prediction purposes, we use 
the full dataset AME2016+ for training,
and carry out  calculations based on a  large set of nuclei 
for which raw theoretical separation energies are not too negative;
this includes all  proton-bound nuclei.
Nuclei with negative experimental separation energies {have not been} used for training.

\subsection{Bayesian model averaging}

When combining several models $\mathcal{M}_k$, the classical literature uses Bayesian posterior weights conditional to the data $y$ given by  \cite{Hoeting1999,Was00,Bernardo1994} 
\begin{equation}
p(\mathcal{M}_k|y)
= \frac{p(y|\mathcal{M}_k)\pi(\mathcal{M}_k)}{\sum_{\ell=1}^K p(y|\mathcal{M}_\ell) \pi(\mathcal{M}_\ell)}, 
\end{equation}
where $\pi(\mathcal{M}_k)$ are the prior weights and $p(y|\mathcal{M}_k)$ are evidence (integrals) obtained by integrating the likelihood over the parameter space. In our study, we shall use uniform priors. 

First, we carry {out} model mixing calculations with  the prior average of the models, i.e., with uniform weights. In the absence of additional information and costly posterior computations, 
the choice of uniform weights is essentially optimal \cite{Bernardo1994}.
This variant is denoted as BMA-0.

Similarly as in Ref.~\cite{Neufcourt2019}, in the context of extrapolations,
we want to select the models with the best predictive power, and avoid overfitting. 
To this end, we evaluate the evidence integrals on new independent data that are  not included in the training of any individual statistical model.
(In contrast, \emph{stricto sensu} BMA would compute the evidence 
based on the whole same dataset as used in the training of each model.) 

In a first variant of BMA calculations (denoted BMA-I), in the spirit of \cite{Neufcourt2019} we  consider 
 simplified Bayesian weights where the evidence is replaced by the posterior probability that each model accounts correctly for the signs of the experimental $Q_{2p}$ and $S_{1p}$ values of the five $2p$ emitters $x_{2p, \rm{known}}$
according to \eqref{eq:2p}. That is, in this variant, the weights are computed
based on the ability of model $\mathcal{M}_k$ to predict the set $x_{2p, \rm{known}}$ as true $2p$ emitters. Here,  the resulting  weights are:
\begin{equation}
w_k({\rm I})  \propto p\left(\mathcal{M}_k | Q_{2p} > 0, S_{1p} > 0 \text{ for } x_{2p, \rm{known}}\right).
\label{eq:w2p}
\end{equation}

In the second variant of BMA calculations (denoted BMA-II),  the evidence $p(y^*|y, \mathcal{M}_k)$ is defined based on the ability of model $\mathcal{M}_k$ to predict known $Q_{2p}$ values of five experimentally known 2$p$ emitters (this set of nuclei is called
$x_{2p, \rm{known}}$$\equiv$$\{$$^{19}$Mg, $^{45}$Fe, $^{48}$Ni, $^{54}$Zn, $^{67}$Kr$\}$ in the following).
The resulting  weights are:
\begin{equation}
w_k({\rm II}) \propto p\left(\mathcal{M}_k | Q_{2p}~\text{of}~ x_{2p, \rm{known}} \right).
\label{eq:e2p}
\end{equation}

Finally, we also consider BMA-III, a trivial version of
BMA-II, consisting in the Gaussian likelihood of $x_{2p, \rm{known}}$ evaluated at the posterior mean and posterior variance values, assuming that these  quantities {\it are independent}.
Considering the
$Q_{2p}$-residuals $y_i$ of the five known $2p$ emitters,   the corresponding  weights are:
\begin{equation}
w_k({\rm III})  \propto \prod_{i \in  x_{2p, \rm{known}}} \frac{1}{\sqrt{2\pi\sigma_i^2}}\,e^{-\frac{1}{2}\left(\frac{y_i}{\sigma_i}\right)^2}.
\label{eq:e2p0}
\end{equation}
Compared to BMA-II, this approximation neglects 
all correlation effects as well as Gaussianity defects 
of the posterior predictions at the five locations
that we meticulously added to the Gaussian process.

\subsection{Model weights computation}

The evidence integrals
$p(y^*|y, \mathcal{M}_k)$ are obtained 
by ``recycling'' Monte-Carlo samples using the estimate
\begin{equation}
w_k=\widehat{p(y^*|y,\mathcal{M}_k)} 
:= \frac{1}{n_{MC}}
\sum_{i} p(y^*|y,\mathcal{M}_k,\theta_k^{(i)}),
\end{equation}
where $\theta_k^{(i)}$ is the $i^{th}$ parameter sample of  $\mathcal{M}_k$.
This is justified by the formula of total probability,
\begin{equation}
p(y^*|y,\mathcal{M}_k)
= 
\int p(y^*|y,\mathcal{M}_k,\theta_k) dP(\theta_k|\mathcal{M}_k,y).
\end{equation}
For each model, this calculation can be efficiently performed in two steps. In the first step, we  compute $q_k^{(i)}:= \log p(y^*|y, \mathcal{M}_k, \theta_k^{(i)})$ 
for each MCMC sample $\theta_k^{(i)}$. In the second step, the weights are obtained as
$$w_k=
e^{q_{max}} 
\frac{1}{n} 
\sum_{i=1}^n e^{q_k^{(i)} - q_{max}}
= \frac{1}{n_{MC}}  
\sum_{i=1}^n p(y^*|y, \mathcal{M}_k, \theta_k^{(i)}).$$
The  testing dataset used to compute the weights overlaps the sets of
even-$N$ and odd-$N$ nuclei. 
In our final  analysis, we have assumed that the statistical models are independent of $N$-parity, meaning that these datasets can be divided as $y^*:=(y_e^*,y_o^*)$.
From the independence of $y_e^*|y_e, \mathcal{M}_k,  \theta_k$ 
and $y_o^*|y_e, \mathcal{M}_k,  \theta_k$ for each model it follows:
\begin{eqnarray}
\log p(y^*|y, \mathcal{M}_k, \theta_k^{(i)})
&=& 
\log p(y_e^*|y_e, \mathcal{M}_k, \theta_{k,e}^{(i)})\nonumber \\
&+&
\log p(y_o^*|y_o, \mathcal{M}_k, \theta_{k,o}^{(i)})
\end{eqnarray}
and the final result is a sum  $q_i = q_i^{(e)}+q_i^{(o)}$.

\begin{figure*}[htb]
\includegraphics[width=0.8\linewidth]{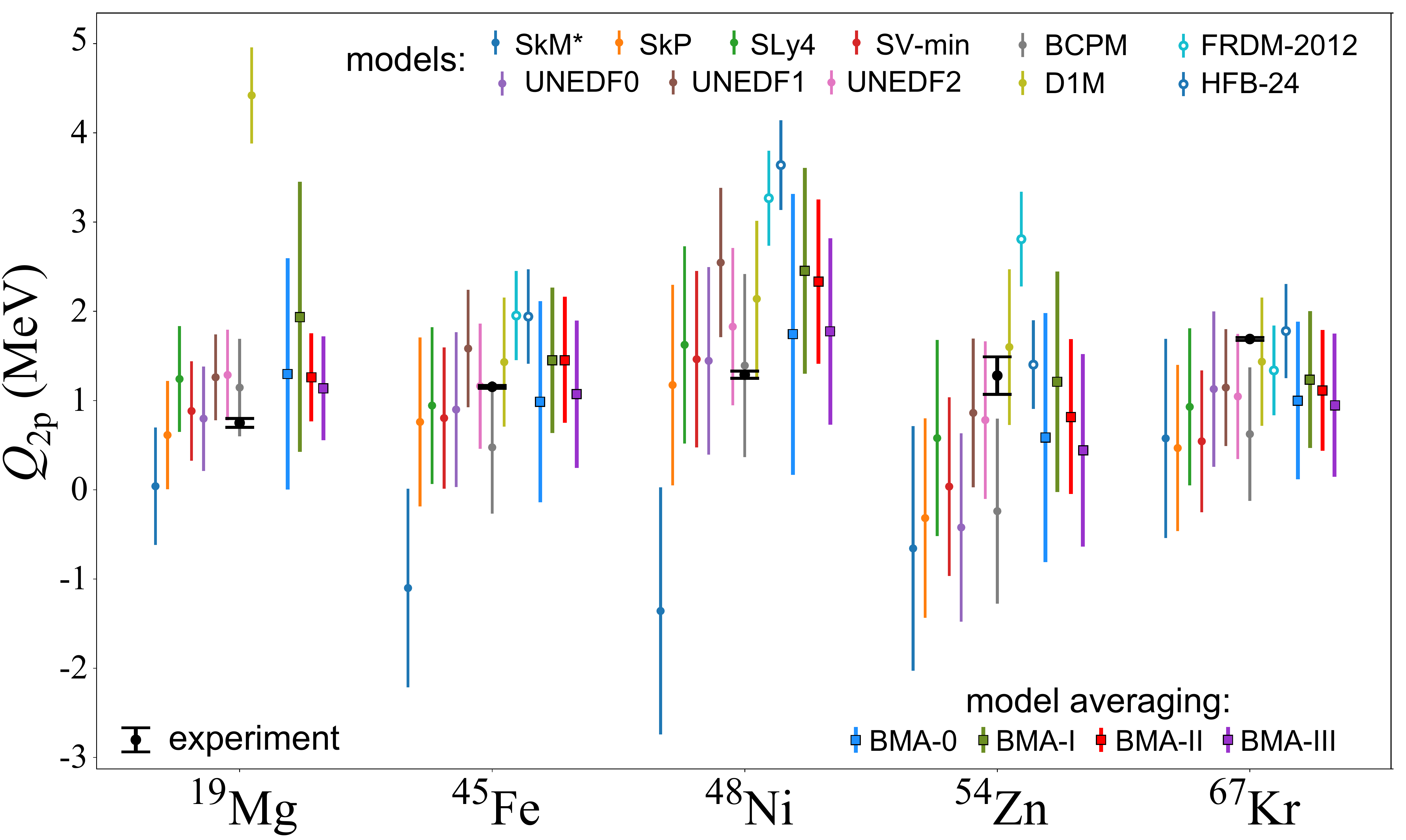}
\caption{$Q_{2p}$ values for the five experimentally known 2$p$ emitters
calculated with the eleven global mass models with statistical correction obtained with GP ($\mu\ne 0$) trained on the AME2016+ dataset. Error bars on theoretical results are one-sigma credible intervals computed with GP extrapolation.
Theoretical results are listed in the following order: Skyrme models SkM$^*$, SkP, SLy4, SV-min, UNEDF0, UNEDF1, and UNEDF2; new models BCPM and D1M; global mass models FRDM-2012 and HFB-24; and model mixing results BMA-0, BMA-I, BMA-II, and BMA-III.
Experimental values are shown for comparison. 
\label{fig:Q2p}
}
\end{figure*}

Note that the assumption of independence is an important \emph{caveat}, and that this calculation would be enhanced by better estimates of the correlations between the systematic errors of $S_{1p}$ and $S_{2p}$.

\section{Results}\label{Results}
\subsection{Model performance and assessment}

Figure~\ref{fig:Q2p} shows {the} comparison between theoretical predictions and experimental data for $^{19}$Mg, $^{45}$Fe, $^{48}$Ni, $^{54}$Zn, and $^{67}$Kr.
We find an overall good agreement with experimental values, up to error bars, for most statistically corrected models. 
It is seen that  individual models, e.g., SkM$^{*}$, HFB-24, and D1M, are behaving  singularly for  some nuclei; this shall impact the model weights used in the following BMA analysis.
The worst performer is the traditional SkM$^*$ model, which practically misses all experimental data points. The best performers are the UNEDF1, UNEDF2, and SLy4  models, which provide the lowest rms deviation from experiment.

Table~\ref{tab:rms} illustrates the global performance of individual models for the measured proton separation energies contained in the AME16-03 testing  dataset.
It is seen that the statistical-model correction brings a significant improvement to the predictions. Namely, in the GP ($\mu=0$) variant, the rms deviation is reduced by
 about $25\%$, {similarly} to our previous studies \cite{Neufcourt2018,Neufcourt2019}, while the $\mu\ne0$ calculation brings another $15\%$ reduction.
The improvement is most significant on the $S_{2p}$ values, and for the models with the largest raw (i.e., statistically-uncorrected)  rms deviations. The improvement for 
FRDM-2012 and HFB-24, optimized to the experimental mass table, is minor, and both variants of GP calculations yield practically identical results.
As in the previous work \cite{Neufcourt2018}, the  rms deviations are quite similar across models following the statistical treatment, which suggests that most of the systematic trends have been captured by our statistical models.

\begin{table*}[!htbp]
  \caption{Model performance. The rms deviations for $S_{1p}$ and $S_{2p}$ (in MeV) for individual models $\mathcal{M}_k,$ and the four BMA variants used, calculated on even-$N$ nuclei. Shown are the uncorrected (raw) and GP-corrected (with the GP mean $\mu=0$ 
 and $\mu\ne 0$) model rms values with respect to the AME16-03 testing dataset. The training dataset used here is AME2003. The raw BMA results  correspond to simple averages of the uncorrected model predictions according to the model weights.
When the BMA weights corresponding to $\mu=0$ and $\mu\neq0$ are different, both raw values are given.
  For compactness, the following abbreviations are used: UNEn=UNEDFn (n=0,1,2) and FRDM=FRDM-2012.
  }  \label{tab:rms}%
    \begin{ruledtabular}
    \begin{tabular}{ccccccccccccc|cccc}
\multicolumn{2}{c}{} & {{SkM*}} & {{SkP}} & {{SLy4}} & {{SV-min}} & {{UNE0}} & {{UNE1}} & {{UNE2}}  & {{BCPM}} & {{D1M}} & {{FRDM}} & {{HFB-24}}& {{BMA-0}} & {BMA-I} & {BMA-II} & {BMA-III} \\
\hline \\[-8pt]
 \multirow{2}{*}{{raw}} & {$S_{1p}:$} & 0.86  & 0.44  & 0.50   & 0.46  & 0.57  & 0.54  & 0.44  & 0.72  & 0.56  & 0.44  & 0.79   & 0.39  & 0.40 & 0.43/0.47 & 0.47/0.43  \\
      & {$S_{2p}$:} & 1.87  & 0.69  & 0.61  & 0.55  & 0.75  & 0.62  & 0.67  & 0.80   & 0.61 & 0.71  & 0.67   & 0.43  & 0.40 & 0.57/0.60 & 0.47/0.42  \\[4pt]
   \multirow{2}{*}{$\mu=0$}& {$S_{1p}$:} & 0.65  & 0.39  & 0.49  & 0.43  & 0.49  & 0.47  & 0.42  & 0.66  & 0.47   & 0.40   & 0.71  & 0.40   & 0.39 & 0.41 & 0.46  \\
     & {$S_{2p}$:} & 1.14  & 0.57  & 0.51  & 0.48  & 0.60   & 0.50   & 0.51 & 0.69  & 0.45  & 0.55  & 0.65   & 0.38  & 0.36 & 0.48 & 0.43 \\[4pt]
 \multirow{2}{*}{$\mu\ne0$} & {$S_{1p}$:} & 0.54  & 0.39  & 0.50   & 0.43  & 0.49  & 0.38  & 0.40  & 0.49  & 0.47    & 0.40   & 0.71 & 0.38  & 0.38 & 0.38 & 0.39  \\
      & {$S_{2p}$:} & 0.76  & 0.44  & 0.50   & 0.43  & 0.60   & 0.39  & 0.42   & 0.64  & 0.45  & 0.55  & 0.65  & 0.36  & 0.35  & 0.37 & 0.34 
  \end{tabular}%
  \end{ruledtabular}
\end{table*}

\subsection{Gaussian process parameters}
\begin{figure}[htb!]
	\centering
		\includegraphics[width=1.0\linewidth]{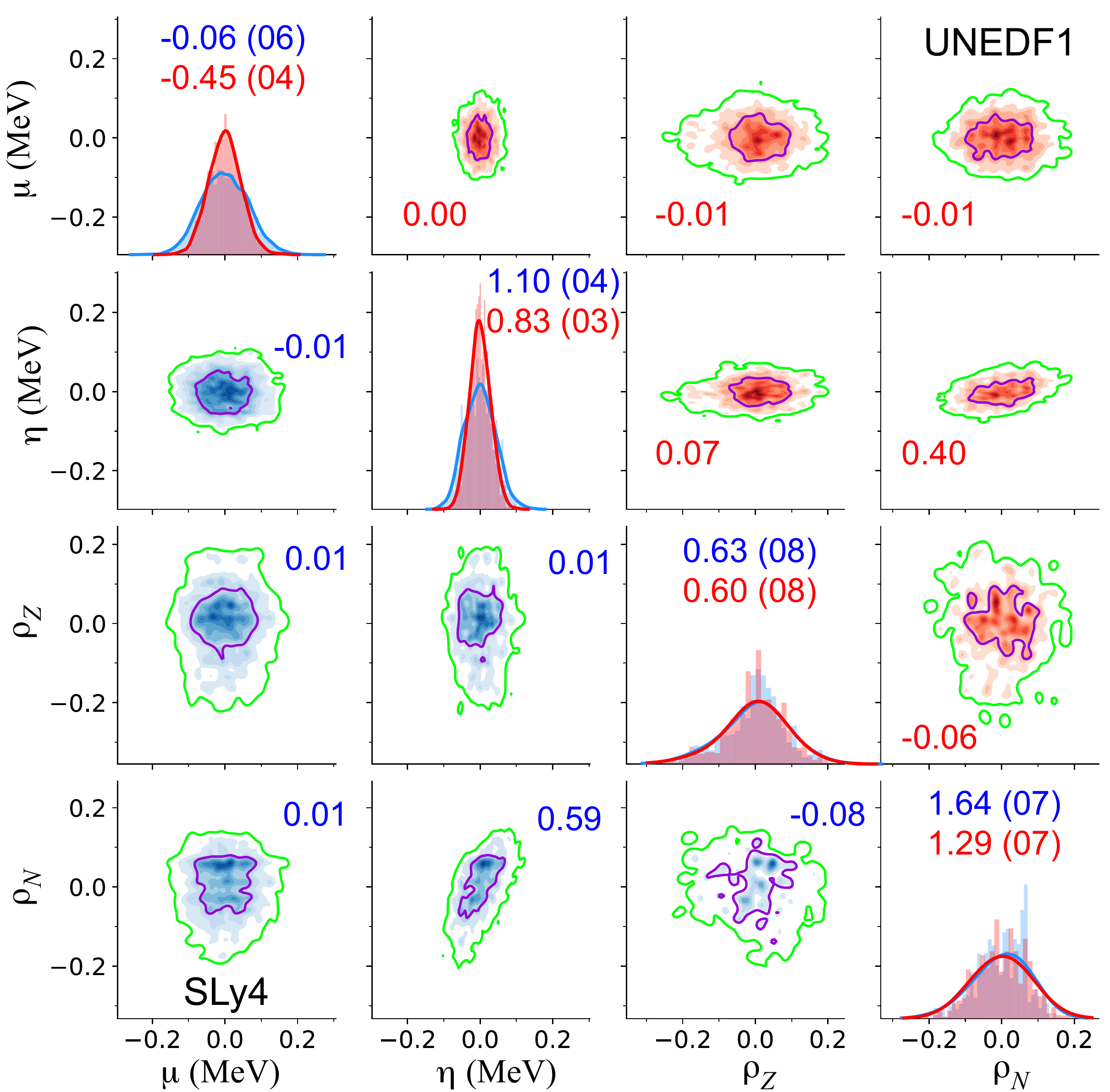}
	\caption{
	Uni- and bivariate distributions of the  50,000  MCMC posterior samples 
	for the 4-dimensional
parameter vector of our Bayesian  statistical model (\ref{GPmodel}-\ref{kernel})
	for UNEDF1 (upper triangle, red) and SLy4 (lower triangle, blue) using the training  $S_{2p}$ dataset of even-even nuclei from AME2003. The parameters are plotted relative to their mean values (corresponding to zero on the plots).
	The diagonal plots show the univariate sample 
	distributions of the GP parameters (histograms) 
	 as well as the KDE estimates (lines). Posterior
mean and standard deviation are indicated by numbers.
The off-diagonal plots show 
	 the KDE estimates of the bivariate posterior distributions 
	 of the GP parameter samples (color map)
	 as well as the 95\% (outer green line) and 50\% (inner purple line)
	  HDRs. The numbers mark the posterior correlation coefficients.
	} \label{fig:pairplot}
\end{figure}
To gain some insight into our statistical model, Fig.~\ref{fig:pairplot} shows the univariate and bivariate distributions of the posterior  samples of $\mu, \eta, \rho_Z$, and $\rho_N$ computed for SLy4 and UNEDF1. Kernel Density Estimates (KDEs) are classically obtained using a Gaussian kernel. Bivariate Highest Density Regions (HDR)s are taken as level lines of the continuous KDE 
\cite{Hyndman1996}.

We can see that the parameters  are both  well constrained and uncorrelated. The only exception is the pair $(\eta,\rho_N)$ for which the correlation coefficient reaches 0.59 for SLy4, which indicates a fairly low  correlation. The dependence of GP parameters on the nuclear model used is also weak, with UNEDF1 producing slightly more localized KDEs. Similar distributions of posterior  samples were obtained  for other models considered.

The posterior mean and standard deviation of the GP parameters
for the two-proton separation energies of even-even nuclei
 are listed in Table~\ref{tab:GPparams}. 
We observe an overall stability of the parameters $\rho_Z$ and $\rho_N$, 
with correlation effects occurring within the range of $\pm(2-3)$ particle numbers. 
Symmetrically to what we have earlier noticed for neutron separation energies
\cite{Neufcourt2018,Neufcourt2019}, we see that 
the correlations effects are here substantially smaller 
in the proton direction than the neutron direction,
consistently with variations in the proton separation energies, 
which are  stronger overall in the $Z$ direction 
than along $N$.

There are more disparities on the parameters $\mu$ and $\eta$, 
which are directly related to the scale of the statistical corrections. 
The parameter $\mu$ is maximal at more than 1 MeV for the model SkM$^*$ and within one standard deviation of zero
 for the more phenomenological models HFB-24 and FRDM-2012. 
The values of $\eta$ are also significantly higher for SkM$^*$ and lower for the two phenomenological models. 
This confirms the common-sense expectation that there is not much left for a GP to capture
 when the nuclear model has already exploited enough of the data structure. 

The scale of the mean and scale parameters $\mu$ and $\eta$ 
appear consistent with the rms improvement in the second row of  Table~\ref{tab:rms}.
As stated above, their largest values were obtained for SkM$^*$, 
for which the rms improvement is also the best.
In this case only 7 masses of spherical nuclei (in addition to other observables) were used in its  optimization process.  Interestingly, the performance of SkP and SLy4 
is outstanding, considering their limited mass input datasets. 
In contrast these parameter values are particularly low 
for HFB-24 and FRDM-2012, 
which were optimized to very large sets of masses.
This confirms that the statistical correction has a balanced autoscaling, and is itself {robust} to overfitting.

A slightly different argument applies to D1M and BCPM, 
for which the  values  of  $\mu$ and $\eta$  are
relatively large in spite of these functionals being optimized 
to a large set of nuclei. It is to be noted, however, that the rotational corrections were added atop the HFB binding  energies during the optimization process of  D1M and BCPM. Since such corrections result in unphysical staggering of separation energies in transitional  nuclei ~\cite{Giuliani2018}, they have been neglected in this work. This
resulted in  increased
values of the mean and scale GP parameters.

The error bars produced for HFB-24 and FRDM-2012, which are scaled by $\eta^2$, 
must be taken with caution, and are certainly undervalued
in the propagation of the training error.
Indeed, the rms deviations are  among the highest on the test data
 (including points more recent than the models) for HFB-24, both before and after statistical correction. 

\begin{table}[htb!]
  \caption{Posterior
mean and standard deviation (in parenthesis) of parameters of our Bayesian  statistical model of $S_{2p}$  for the nuclear mass models used in this study. $\mu$ and $\eta$  are in  MeV  and  $\rho_Z$ and  $\rho_N$ are dimensionless.
The last column (\#m) gives the number of nuclear masses used in nuclear model optimization. For SkM$^*$, SkP, and SLy4 only spherical masses were used.}
    \begin{ruledtabular}
    \begin{tabular}{lccccc}
     Model     & $\mu$ & $\eta$ & $\rho_Z$ & $\rho_N$ & $\#m$ \\ \hline
    \\[-8pt]
    SkM$^*$  & 1.33(0.09)  & 1.38(0.05)  & 0.71(0.10)  & 1.90(0.05) & 7\\
    SkP   & 0.38(0.07)  & 1.11(0.04)  & 0.65(0.09)  & 1.71(0.07) &  2\\
    SLy4  & -0.06(0.06)  & 1.10(0.04)  & 0.63(0.08)  & 1.64(0.07) & 5 \\
    SV-min & 0.24(0.06)  & 1.01(0.04)  & 0.61(0.08)  & 1.53(0.07) & 64 \\
    UNE0 & 0.07(0.06)  & 1.05(0.04)  & 0.64(0.09)  & 1.62(0.06) & 72 \\
    UNE1 & -0.45(0.04)  & 0.83(0.03)  & 0.60(0.08)  & 1.29(0.07) & 75 \\
    UNE2 & -0.38(0.05)  & 0.88(0.03)  & 0.61(0.08)  & 1.37(0.07) & 76 \\
    BCPM  & 0.21(0.06)  & 1.03(0.04)  & 0.76(0.11)  & 1.47(0.06) & 579\\
    D1M   & 0.15(0.05)  & 0.87(0.03)  & 0.66(0.09)  & 1.55(0.05) & 2149\\[4pt]
    HFB-24 & 0.01(0.03)  & 0.50(0.02)  & 0.64(0.09)  & 1.29(0.06) & 729\\
    FRDM & 0.04(0.03)  & 0.53(0.02)  & 0.67(0.09)  & 1.78(0.05) & 2149\\
 \end{tabular}
\end{ruledtabular}
  \label{tab:GPparams}%
\end{table}%

\begin{table*}[!htb]
  \caption{Model posterior weights obtained in {the} variants BMA-I (\ref{eq:w2p}), BMA-II  (\ref{eq:e2p}),
  and BMA-III (\ref{eq:e2p0}) of our BMA calculations
based on the AME16+ (top)  and AME03 (bottom) training datasets.
 }
  \label{tab:weights}%
  \begin{ruledtabular}
    \begin{tabular}{lcccccccccccc}
  BMA variant & {SkM*} & {SkP} & {SLy4} & {SV-min} & {UNE0} & {UNE1} & {UNE2} & {BCPM} & {D1M} & {FRDM} & {HF-B24} \\ \hline
    \\[-8pt]
BMA-I:  & 0.00  & 0.03  & 0.08  & 0.05  & 0.04  & 0.14  & 0.12  & 0.04  & 0.16 & 0.17  & 0.17 \\ 
BMA-II: & 0.00  & 0.00  & 0.00  & 0.00  & 0.01  & 0.71  & 0.27  & 0.00  & 0.00 & 0.00  & 0.00 \\
BMA-III: & 0.00  & 0.05  & 0.17  & 0.10  & 0.11  & 0.16  & 0.35  & 0.05  & 0.00 & 0.00  & 0.00 \\
\hline    \\[-8pt]
BMA-I: & 0.00 & 0.02 & 0.07 & 0.04 & 0.04 & 0.14 & 0.12 & 0.04 & 0.16  & 0.19 & 0.19 \\
BMA-II & 0.00 & 0.00 & 0.00 & 0.00 & 0.01 & 0.51 & 0.47 & 0.00 & 0.00 & 0.00 & 0.00 \\
BMA-III: & 0.00 & 0.04  & 0.15 & 0.08 & 0.11 & 0.19  & 0.38 & 0.06 & 0.00 & 0.00 & 0.00 
 \end{tabular}
\end{ruledtabular}
\end{table*}%

\subsection{Model mixing performance and assessment}

The model weights obtained in the BMA variants  used  are listed in Table~\ref{tab:weights}.  The  weights of BMA-I are rather uniformly distributed, with the highest values for UNEDF1, UNEDF2, D1M, FRDM-2012, and HFB-24; the weight 0 for SkM$^*$ is due to the fact that it misses completely the sign of the second and third testing points (see Fig.~\ref{fig:Q2p}), which is eliminatory. In the case of the likelihood-based  BMA-II, 
only the Skyrme models UNEDF1 and UNEDF2 practically contribute.
This can be explained by the concavity of the evidence with respect to the data, which heavily penalizes large deviations at single locations.

Figure~\ref{fig:Q2p} shows the results of BMA  for the five true $2p$ emitters considered. It is encouraging to see that the BMA predictions agree well with experimental values, i.e., their standard error bars overlap.
As  expected theoretically \cite{Kej19},  the BMA results should achieve the lowest rms deviations among all models. This is confirmed in Table~\ref{tab:rms}, which shows that the rms deviations for $S_{1p}$ and $S_{2p}$ obtained in all BMA calculations are indeed below those of individual nuclear mass models.

 We see in Table~\ref{tab:rms} that {all BMA variants} perform  very similarly overall:
 the rms deviations are   around  0.38\,keV for $S_{1p}$ and 0.36\,MeV for $S_{2p}$.
It
may be thus tempting to associate these values with the precision limit that current
EDFs can achieve in the description of proton separation energies. 
 Interestingly, the simple models  BMA-0 and BMA-III  yield rms deviations comparable, if not lower, to the other elaborate variants.
This is not entirely surprising: the posterior weights favor by construction the statistically corrected models best fitted at locations $y^*$.
Indeed, some of the nuclear models employed, such as HFB-24 and FRDM-2012 are less robust to out-of-sample testing data, which can possibly be linked to some overfitting.

Figure~\ref{fig:Q2p} shows that the posterior means of BMA-III are -- by construction -- close to the experimental $Q_{2p}$ values. The disadvantage of
this approximation to BMA-II is that it neglects all correlation effects and long range dependencies meticulously added to the GP. Consequently, 
the differences between the BMA-II and BMA-III  weights can be large for some models, cf. the weights for SLy4 and UNEDF1 in Table~\ref{tab:weights}. This result can be traced back to  non-zero covariances between the five locations and Gaussianity defaults. First, the (posterior) covariances between the different $Q_{2p}$ values are non-zero in the actual samples, but assumed uniformly zero in BMA-III. 
Second, the evidence calculation in BMA-II includes 
an integration over the parameter space, which is absent in BMA-III.
For instance, the parameter $\rho_N$ is 27\% larger for SLy4 than in UNEDF1, and
 this means that one more neighbor in the neutron direction is roughly included in the GP calculations based on SLy4. While the parameter difference is not dramatic, it does significantly impact {both} the predictions and model weight estimates. 

\begin{figure*}[htb!]
\includegraphics[width=0.9\linewidth]{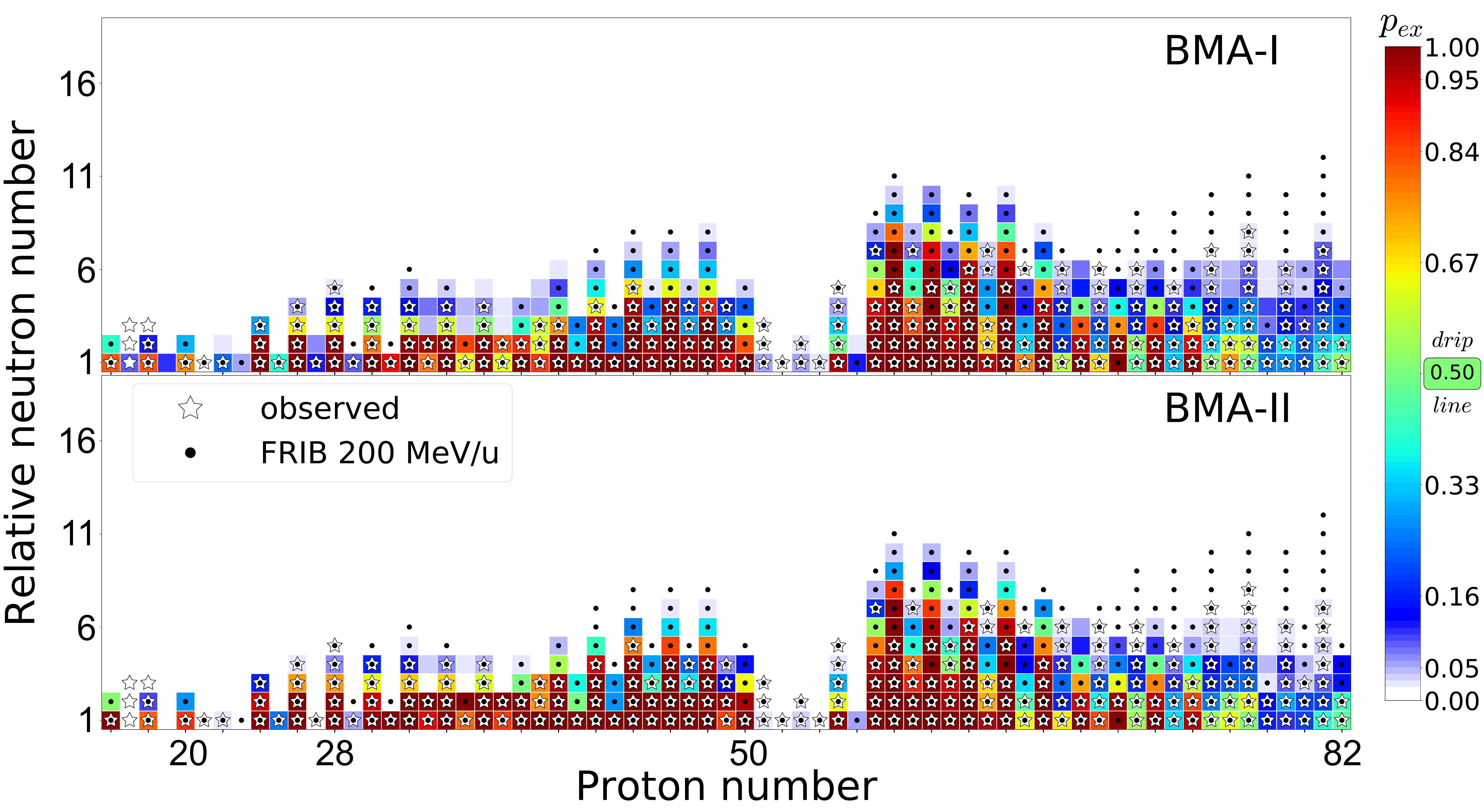}
\caption{The quantified nuclear binding-energy landscape in the proton rich region obtained in BMA-I (top) and BMA-II (bottom) model averaging calculations.
The color marks the probability $p_{\rm ex}$  that  these nuclei are bound with respect to  proton decay. 
For each proton number, $p_{\rm ex}$ is shown  along the isotopic chain versus
 the relative neutron number $N_0(Z)-N$, where
 $N_0(Z)$, listed in Tables~\ref{tab:reference} and \ref{tab:reference1}, 
 is the neutron number of the lightest proton-bound isotope
 for which an experimental one- or two-proton separation energy value is available.
The domain of nuclei that have been experimentally observed 
(both proton-bound and proton-unbound)
is marked by open stars; those
within  FRIB's experimental reach are marked by dots. See text for details.
\label{fig:landscape}
}
\end{figure*}

\subsection{Predictions of BMA calculations: quantified landscape of proton-rich nuclei}

The quantified nuclear binding-energy landscape for  proton rich nuclei, predicted in BMA-I and BMA-II,  is displayed in   Fig.~\ref{fig:landscape}. To facilitate the presentation, the information for each isotope is given  relative to the neutron number 
$N_0$ of the lightest proton-bound isotope
 for which an experimental one- or two-proton separation energy value is available.
In analogy with Ref.~\cite{Neufcourt2019} we show the probability  $p_{\rm ex}$  that  a given isotope is proton-bound, i.e., that  $S_{2p}>0$ for even-$Z$ nuclei and  $S_{1p}>0$ for odd-$Z$ nuclei. Formally, in the Bayesian language, this quantity can be defined as:
\begin{equation}\label{pex}
p_{\rm ex}:=p(S_{1p/2p}^* > 0 | S_{1p/2p}).
\end{equation}
The proton drip line corresponds to $p_{\rm ex}=0.5$.
The reference values of  $N_0(Z)$ are listed in Tables~\ref{tab:reference} (for even-$Z$ nuclei) and 
\ref{tab:reference1} (for odd-$Z$ nuclei), together with the range of observed nuclei and proton drip-line nuclei.

\begin{figure}[htb!]
\includegraphics[width=1.0\linewidth]{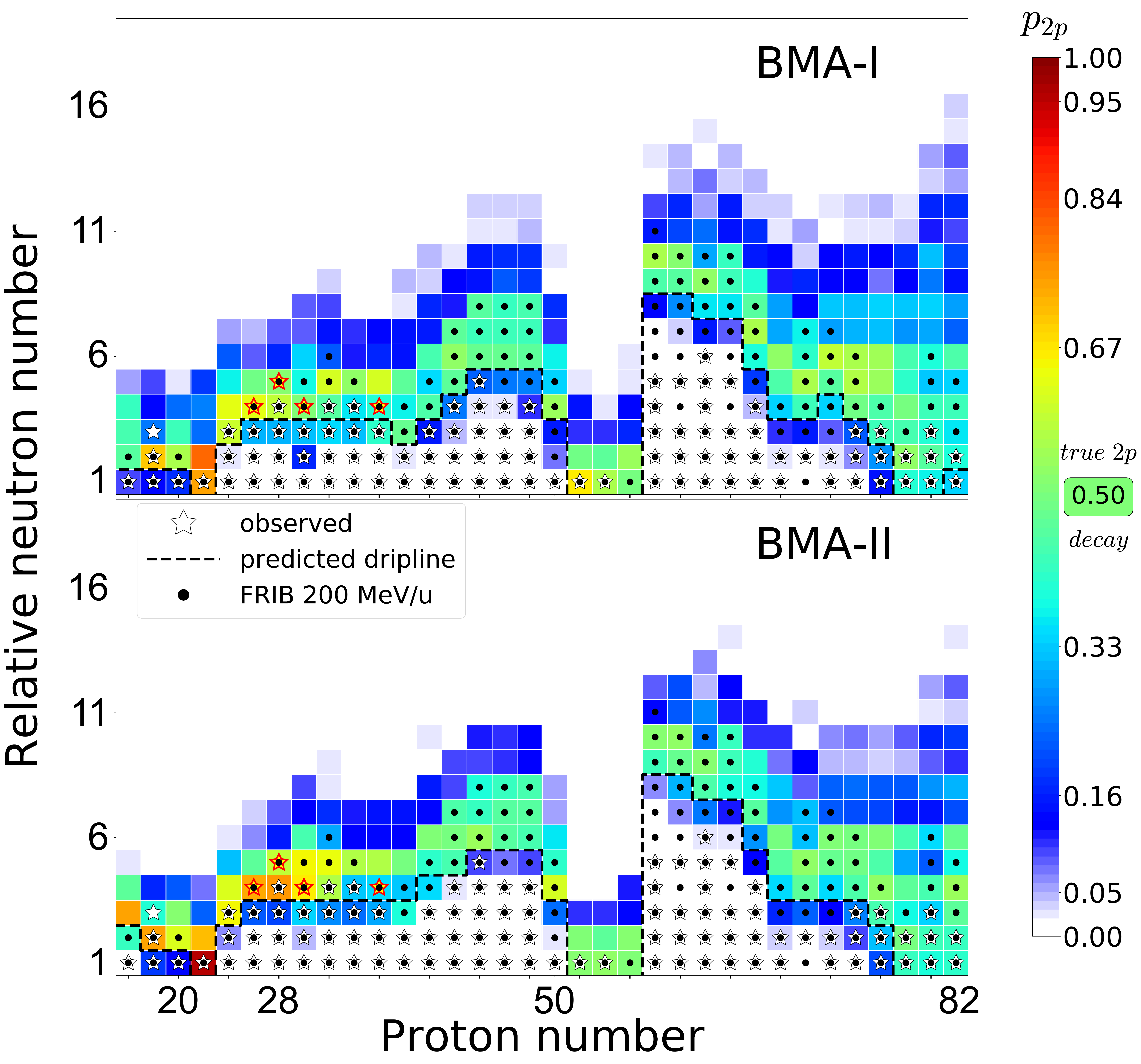}
\caption{Probability of true $2p$ emission for the even-$Z$ proton-rich isotopes.
The color gives the posterior probability of $2p$ emission, i.e.,  that $S_{2p} < 0$ and $S_{1p} > 0$, according to the posterior average models.
For each proton number, shown  is the relative neutron number $N_0(Z)-N$, where
 $N_0(Z)$, listed in Tables~\ref{tab:reference} and \ref{tab:reference1}, is the neutron number of the lightest proton-bound isotope for which an experimental two-proton separation energy value is available.
The dotted line represents the predicted drip line  (corresponding to $p_{ex} = 0.5$).
The domain of nuclei which have been experimentally observed is marked by stars (the experimentally observed $2p$ emitters $^{45}$Fe, $^{48}$Ni,  $^{54}$Zn,
and  $^{67}$Kr are indicated by closed stars); those
within  FRIB's experimental reach are marked by dots. See text for details.
\label{fig:p2p}
}
\end{figure}

Figure~\ref{fig:landscape} and Tables \ref{tab:reference} and \ref{tab:reference1} also show
the isotopes that will be accessible at the future  Facility for Rare Isotope Beams (FRIB) \cite{Glasmacher17,Sherrill}. FRIB will accelerate ion species up to $^{238}$U with energies of no less than 200 MeV/u (beam power up to 400 kW). 
The FRIB production rates have been  calculated with the \lisepp code \cite{LISE} designed to predict intensity and purity for future experiments using rare beams with in-flight separators. 
The EPAX2.15 cross-section systematics \cite{Summerer} and the \lisepp 3EER Abrasion-Fission model \cite{LISEpp,LISEpp2} were used to calculate production cross-sections for projectile fragmentation and fission reactions correspondingly. The multi-step reactions in thick targets were taken into account. In this process, the projectile undergoes a series of successive reactions until the fragment of interest is produced. FRIB Rates and details of their calculations are available online \cite{fribrates}. In our estimates, we assumed  the  experimental limit for the confirmation of existence of an isotope to be 1 event/2.5 days.

\begingroup
\squeezetable
\begin{table}[htb!]
  \caption{Reference table to Fig.~\ref{fig:landscape}: even-$Z$ elements.
  For each atomic element with even-$Z$ shown are: the neutron number $N_0$ of the lightest isotope for which an experimental one- or two-proton separation energy value is available; 
  the neutron number $N_{\rm obs}$ of the lightest isotope observed; 
  the neutron number $N_{\rm drip}$ of the predicted drip line isotope in BMA-I; and the neutron number $N_{\rm FRIB}$ marking the reach of FRIB.
  }
    \begin{ruledtabular}
    \begin{tabular}{rlrccc}
        \multicolumn{2}{c}{$Z$}  & $N_0$ & $N_{\rm obs}$  &  $N_{\rm drip}$ &  $N_{\rm FRIB}$\\ \hline 
            \\[-8pt]
    16    & S     & 12    & 11    & 11    & 10 \\
    18    & Ar    & 14    & 11    & 13    & 12 \\
    20    & Ca    & 16    & 15    & 15    & 14 \\
    22    & Ti    & 18    & 17    & 18    & 17 \\
    24    & Cr    & 21    & 18    & 19    & 18 \\
    26    & Fe    & 23    & 19    & 20    & 19 \\
    28    & Ni    & 25    & 20    & 22    & 20 \\
    30    & Zn    & 28    & 24    & 25    & 23 \\
    32    & Ge    & 31    & 27    & 28    & 25 \\
    34    & Se    & 33    & 29    & 30    & 28 \\
    36    & Kr    & 35    & 31    & 32    & 31 \\
    38    & Sr    & 37    & 35    & 35    & 33 \\
    40    & Zr    & 40    & 37    & 37    & 35 \\
    42    & Mo    & 43    & 39    & 39    & 36 \\
    44    & Ru    & 46    & 41    & 41    & 38 \\
    46    & Pd    & 48    & 44    & 43    & 40 \\
    48    & Cd    & 50    & 46    & 45    & 42 \\
    50    & Sn    & 50    & 49    & 47    & 45 \\
    52    & Te    & 53    & 52    & 53    & 52 \\
    54    & Xe    & 55    & 54    & 55    & 54 \\
    56    & Ba    & 58    & 58    & 58    & 57 \\
    58    & Ce    & 68    & 63    & 60    & 57 \\
    60    & Nd    & 70    & 65    & 62    & 60 \\
    62    & Sm    & 73    & 67    & 66    & 63 \\
    64    & Gd    & 76    & 71    & 69    & 66 \\
    66    & Dy    & 77    & 73    & 72    & 69 \\
    68    & Er    & 78    & 76    & 75    & 74 \\
    70    & Yb    & 81    & 79    & 78    & 74 \\
    72    & Hf    & 84    & 82    & 80    & 77 \\
    74    & W     & 86    & 83    & 83    & 80 \\
    76    & Os    & 88    & 85    & 86    & 84 \\
    78    & Pt    & 90    & 88    & 90    & 87 \\
    80    & Hg    & 94    & 91    & 94    & 88 \\
    82    & Pb    & 98    & 96    & 97    & 93 \\
    \end{tabular}%
  \label{tab:reference}%
\end{ruledtabular}
\end{table}
\endgroup

\begingroup
\squeezetable
\begin{table}[htb!]
  \caption{Reference table to Fig.~\ref{fig:landscape}: odd-$Z$ elements.
  Similar as in Table~\ref{tab:reference} but for odd-$Z$ isotopes.
  }
    \begin{ruledtabular}
    \begin{tabular}{rlrccc}
        \multicolumn{2}{c}{$Z$}  & $N_0$ & $N_{\rm obs}$  &  $N_{\rm drip}$ &  $N_{\rm FRIB}$\\ \hline 
            \\[-8pt]
   17    & Cl    & 14    & 11    & 14    & 14 \\
    19    & K     & 16    & 16    & 16    & 16 \\
    21    & Sc    & 19    & 18    & 19    & 18 \\
    23    & V     & 20    & 20    & 20    & 19 \\
    25    & Mn    & 22    & 21    & 22    & 21 \\
    27    & Co    & 24    & 23    & 24    & 23 \\
    29    & Cu    & 27    & 26    & 27    & 25 \\
    31    & Ga    & 30    & 29    & 29    & 28 \\
    33    & As    & 33    & 31    & 31    & 31 \\
    35    & Br    & 35    & 34    & 33    & 33 \\
    37    & Rb    & 37    & 35    & 35    & 35 \\
    39    & Y     & 40    & 37    & 37    & 37 \\
    41    & Nb    & 42    & 41    & 41    & 39 \\
    43    & Tc    & 44    & 43    & 43    & 40 \\
    45    & Rh    & 47    & 44    & 45    & 42 \\
    47    & Ag    & 49    & 45    & 47    & 44 \\
    49    & In    & 51    & 47    & 49    & 47 \\
    51    & Sb    & 55    & 52    & 55    & 52 \\
    53    & I     & 57    & 55    & 57    & 55 \\
    55    & Cs    & 62    & 57    & 61    & 57 \\
    57    & La    & 67    & 60    & 61    & 58 \\
    59    & Pr    & 69    & 62    & 65    & 61 \\
    61    & Pm    & 72    & 67    & 68    & 64 \\
    63    & Eu    & 74    & 67    & 71    & 67 \\
    65    & Tb    & 76    & 70    & 75    & 69 \\
    67    & Ho    & 79    & 73    & 78    & 72 \\
    69    & Tm    & 82    & 76    & 81    & 75 \\
    71    & Lu    & 85    & 79    & 83    & 76 \\
    73    & Ta    & 87    & 82    & 87    & 78 \\
    75    & Re    & 91    & 84    & 90    & 81 \\
    77    & Ir    & 95    & 87    & 93    & 84 \\
    79    & Au    & 97    & 91    & 97    & 87 \\
    81    & Tl    & 102   & 95    & 102   & 90 \\
    \end{tabular}%
  \label{tab:reference1}%
\end{ruledtabular}
\end{table}
\endgroup

In general, the drip-line predictions of BMA-I and BMA-II are very similar; some differences can be see for the elements just below Pb, with BMA-I calculating slightly  higher values of $p_{\rm ex}$. In this region of nuclei, FRIB is expected to have a particularly high discovery potential: by exploring the vast territory of proton-unstable isotopes, it will extend the domain of  known nuclei considerably.

The territory of the true $2p$ emitters predicted in our BMA calculations is shown in Fig.~\ref{fig:p2p}. Here, the quantity of interest is the posterior probability $p_{2p}$  that $S_{2p} < 0$ and $S_{1p} > 0$:
\begin{equation}\label{p2p}
p_{2p}:=p(S_{2p}^* < 0 \cap S_{1p}^* > 0| S_{1p/2p}),
\end{equation}
see Eq.~\eqref{eq:2p}. Again,  BMA-I and BMA-II predictions are close. The isotopes that are potential candidates for $2p$ radioactivity lie in a band corresponding to $p_{2p}\geq 0.5$.
Most of those isotopes are within the range of FRIB.

\subsection{Two-proton radioactivity: lifetime considerations}

In the previous sections, the discussion of proton radioactivity was solely based on 
 energy arguments. However, the energy relations do not tell the full story. Indeed, the proton decays corresponding to very large $Q_{1p/2p}$-values are going to be too fast to be observed. If the $Q_{1p/2p}$-values are very low, the proton-decay rate is going to be negligible compared to other decay modes, such as $\beta$ or $\alpha$ decays. 
When it comes to $2p$ decay, {the practical range of lifetimes 
is}~\cite{Olsen2p}: 
\begin{equation} \label{eq:T}
10^{-7}\,{\rm s} < T_{2p} < 10^{-1}\,{\rm s}.
\end{equation}
The lower bound of 100 ns corresponds to the typical sensitivity limit of in-flight, projectile-fragmentation techniques. The upper bound of 100 ms ensures that {the} $2p$ decay will not be dominated by  $\beta$ decay.

To get an order-of-magnitude estimate of $2p$ lifetimes, we used the simple  semiclassical Wentzel-Kramers-Brillouin (WKB)  approximation, and assumed  a diproton decay with $\ell=0$. For details of {the} WKB calculations, see~\cite{Olsen2p,Witek1996}.
The value of the proton overlap  $\mathcal{O}^2$ has been fitted to match the measured lifetimes of $^{19}$Mg, $^{45}$Fe, $^{48}$Ni, $^{54}$Zn, {yielding $\mathcal{O}^2=0.0011$}.
Our WKB approach agrees very well with the semiclassical effective liquid drop model analysis of Ref.~\cite{Goncalves2017}.

\begin{figure*}[htb!]
\includegraphics[width=1.0\linewidth]{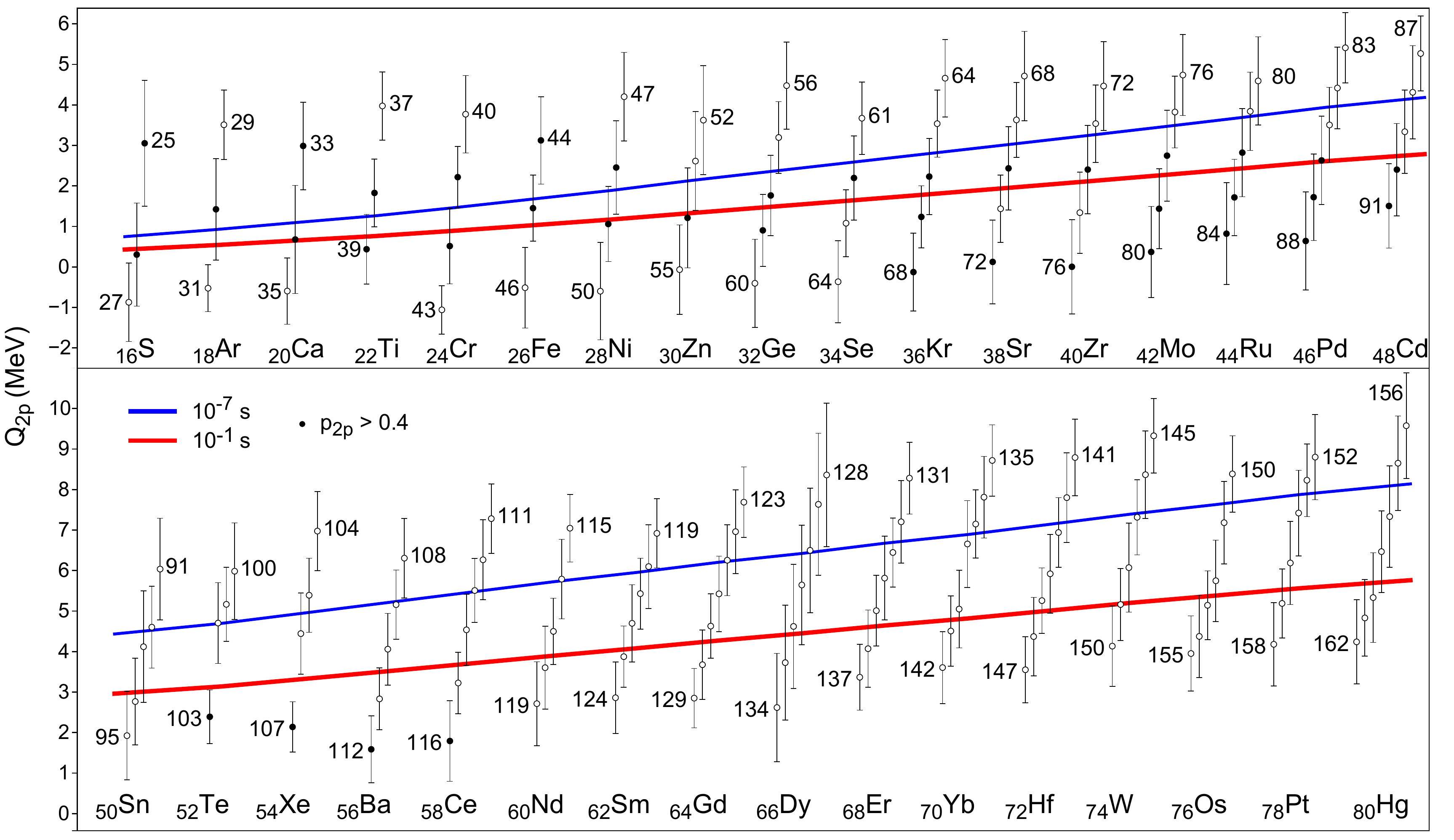}
\caption{$Q_{2p}$ values predicted in BMA-I for even-even isotopes  with $16 \le Z \le 80$. The thick solid lines mark the lifetime range (\ref{eq:T}).
The mass numbers of selected isotopes are shown.
The nuclei with the probability $p_{2p}>0.4$ are indicated by dots. Here, we used this value of $p_{2p}$ rather than $p_{2p}>0.5$  because the criterion (\ref{p2p}) of the true $2p$ emission is slightly more  restrictive  than the energy criterion previously adopted in Ref. \cite{Olsen13}.
\label{fig:lifetimes}
}
\end{figure*}

Figure~\ref{fig:lifetimes} shows the $Q_{2p}$ values predicted in BMA-I together with the lifetime range (\ref{eq:T}).
It is important to note that the  uncertainties on the predicted values of  $Q_{2p}$  usually correspond to several {decades} of the $2p$-decay lifetime.
As seen in Fig.~\ref{fig:lifetimes},
the known $2p$ emitters 
$^{45}$Fe, $^{48}$Ni, $^{54}$Zn, and  $^{67}$Kr  consistently fall within the
lifetime range (\ref{eq:T}). The most promising other candidates for the true $2p$ radioactivity are:
${}^{30}\text{Ar}$, ${}^{34}\text{Ca}$, ${}^{39}\text{Ti}$, ${}^{42}\text{Cr}$, 
${}^{58}\text{Ge}$, ${}^{62}\text{Se}$, ${}^{66}\text{Kr}$, ${}^{70}\text{Sr}$,
${}^{74}\text{Zr}$, ${}^{78}\text{Mo}$, ${}^{82}\text{Ru}$, ${}^{86}\text{Pd}$, ${}^{90}\text{Cd}$, and ${}^{103}\text{Te}$.

For nuclei with $Z\ge 54$ that are within the lifetime range (\ref{eq:T}), our calculations predict  $p_{2p}<0.4$, i.e., low probability {of} true $2p$ emission.  Indeed, for heavy nuclei, because of the large Coulomb barriers, the condition of   $p_{2p}>0.4$ corresponds to low $Q_{2p}$ values and very long lifetimes resulting in  small $2p$ widths. 
According to the results shown in  Figs.~\ref{fig:p2p} and \ref{fig:lifetimes},
such a situation is expected in, e.g.,  ${}^{107}\text{Xe}$, 
${}^{112}\text{Ba}$, ${}^{116}\text{Ce}$,
 ${}^{120}\text{Nd}$, ${}^{126}\text{Sm}$, ${}^{136}\text{Dy}$, ${}^{140}\text{Er}$, ${}^{146}\text{Yb}$, ${}^{150}\text{Hf}$, ${}^{154}\text{W}$, ${}^{155}\text{W}$, ${}^{158}\text{Os}$, ${}^{159}\text{Os}$, and ${}^{166}\text{Pt}$. 

It is also to be noted that 
many of the extremely proton-rich nuclei in Fig.~\ref{fig:lifetimes} with small $p_{2p}$ values, such as $^{131,132}$Dy, $^{134,135}$Er  and ${}^{144,145}$Hf, are excellent candidates for the {\it sequential} emission of two protons ($pp$)  \cite{Olsen13}.

Our BMA findings are fairly consistent with predictions of other papers.
The nuclei $^{39}$Ti ($p_{2p}= 0.74$) and  $^{42}$Cr ($p_{2p}= 0.60$)  are expected to be excellent $2p$-decay candidates according to the phenomenological analysis based on the
modified Kelson-Garvey
mass relations \cite{Tian2013} and shell model analysis \cite{Brown1991}.
As discussed in  \cite{Dossat2007}, ${}^{39}$Ti  primarily decays by beta disintegration. This is not inconsistent with the low  $Q_{2p}$ value predicted in BMA-I. 
Other $2p$-decay candidates predicted by BMA-I  discussed in the literature include: $^{26}$S, $^{29-31}$Ar
\cite{Gigorenko2018},
${}^{34}$Ca \cite{Goncalves2017},
$^{58}$Ge, $^{62}$Se and $^{66}$Kr \cite{Grigorenko2003}. 
The nucleus $^{103}$Te, which  has been predicted \cite{Olsen13} to exhibit a competition between alpha decay and  $2p$ radioactivity, is expected to have 
$T_{2p} > 0.1$\,s. In ${}^{145}$Hf, alpha decay is predicted \cite{Olsen13}  to compete with sequential $pp$ emission.

\section{Conclusions}\label{Conclusions}

In this paper, we employed the Bayesian Model Averaging framework  to quantify the proton-stability of the nucleus. To this end, we  introduced the  probability $p_{\rm ex}$, which is the Bayesian posterior probability that the one- or two-proton  separation energy of a nucleus is positive. We also evaluated the posterior probability $p_{2p}$ that a nucleus is a true $2p$ emitter. 

We demonstrate that the statistical-model correction improves predictions significantly. 
Overall, for the
testing dataset AME16-03, the rms deviation from
experimental $S_{2p}$ values is in the 400-600 keV range in
the GP ($\mu\ne 0$) variant across theoretical models employed
in this study. 
This is consistent with the previous analysis of  $S_{2n}$ values \cite{Neufcourt2018} and
indicates
that our GP model 
captures a significant part of the systematics and brings a sound refinement to the nuclear theory models. 

Following the model averaging, the  rms deviations from experiment for proton separation energies  are surprisingly similar for all the BMA variants used: they are around 360\,keV for $S_{2p}$ and 380\,keV for $S_{1p}$. This result suggests that the further rms reduction  cannot probably be expected without   dramatic  improvements of fidelity of nuclear mass models.

In general, our results are fairly consistent with the current experimental data on the proton drip line position  and the appearance of $2p$ radioactivity. 
Our calculations suggest that no true $2p$ emission is expected in the  lifetime range (\ref{eq:T}) above $Z=54$. 

In contrast to an increasing number of studies applying directly statistical models to nuclear physics observables \cite{Jiang2019,Yoshida2019,Wang2019,Regnier2019}, the sound nuclear physics model underneath our statistical emulators, trained on the residuals, guarantees that our predictions are globally consistent with known physics. The GP parameters are well constrained with relatively weak posterior variances and correlations, and the additional mean parameter brings another reduction of rms deviation. The GP could potentially be refined in future studies by including additional degrees of freedom to describe different regions of the nuclear domain.

As we emphasized in our previous studies \cite{Neufcourt2018,Neufcourt2019}, statistics is no magics, and no statistical model can do more than reproducing patterns found in the existing data (in our case: model residuals). In this context, far extrapolations must rely on quality nuclear modeling.
We believe that the key in our approach is the evaluation of posterior predictive distributions (instead of predicted values) where the mean value itself is of less importance compared to credibility intervals. 

In the course of this study, it has appeared that a non-zero value of the GP mean prediction $\mu$ allows to reproduce more consistently the extrapolative data (kept away during the training). In our opinion,  the GP extension to  nonzero $\mu$ is more reasonable than hypothesizing a more elaborate tail model, which - if not substantiated by physics - would  either be speculative or lead to overfitting. 

We  proposed Bayesian average models obtained by using several variants of the BMA weights; the weights are calibrated on independent test data, and thus directly related to the extrapolative power of each model. While we observe significant variations in the weights obtained {for} different BMA variants, all average models achieve a lower rms deviation than individual model constituents. This validates empirically the essence of the recent BMA analysis \cite{Kej19} where it has been established that the BMA estimator achieves the lowest posterior variance among all models and all model combinations. 
This result suggests that even the simplest uniform model mixing carried out in several previous studies involving different quantities \cite{Erl12a,Kortelainen13,Agbemava2014,Olsen19,Neufcourt2019} can provide a very valuable information.

It may appear statistically disappointing that the BMA-0, BMA-I, and  BMA-III variants achieve the  better testing rms performance as compared to  the most sophisticated BMA-II method. 
Surely the comparison of the BMA weights (BMA-II) with their counterparts obtained from model conditional likelihoods taken at posterior mean values (BMA-III) highlights the Gaussianity defects and correlations, hidden in a standard analysis of rms deviations and error bands.  
The philosophy of using model mixing to attenuate the individual defects of individual models makes it desirable to include a greater diversity of models, and our results suggest that difficulties are to be expected  when none of the proposed statistical models gives an accurate description of the  covariance structure between data points.

In the context of BMA-based extrapolations, we wish to emphasize that nothing can be stated with certainty in  the domain where no experimental data are available.  However it is clear that, on our testing dataset (which corresponds to the outer boundary of the current experimental knowledge, and was not used for training) BMA outperforms every single model (see Table~\ref{tab:rms}). Again, based on general considerations \cite{Kej19}, BMA should on average outperform individual models.

The extrapolation outcomes discussed in this study will be tested by  experimental data from rare-isotope facilities.   As illuminated by our Bayesian  analysis,  experimental discoveries of new proton-rich nuclides   will  be crucial for delineating the detailed behavior  of the nuclear mass surface in the vast unexplored region of the nuclear landscape beyond the proton drip lines.


\begin{acknowledgments}
Computational resources for statistical simulations 
were provided to L.N.
by the Institute for Cyber-Enabled Research at Michigan State University 
as well as Research Credits awarded by Google Cloud Platform. 
L.N. also thanks Vojtech Kejzlar for sharing an efficient code 
to compute evidence integrals. This material is based upon work supported by the U.S.\ Department of Energy, Office of Science, Office of Nuclear Physics under  award numbers DE-SC0013365 (Michigan State University), DE-SC0018083 (NUCLEI SciDAC-4 collaboration), and  DOE-NA0003885
(NNSA, the Stewardship Science Academic
Alliances program).
\end{acknowledgments}


%

\end{document}